\documentclass{article}

\usepackage{PRIMEarxiv}

\usepackage[utf8]{inputenc} 
\usepackage[T1]{fontenc}    
\usepackage{hyperref}       
\usepackage{url}            
\usepackage{booktabs}       
\usepackage{amsfonts}       
\usepackage{nicefrac}       
\usepackage{microtype}      
\usepackage{lipsum}
\usepackage{fancyhdr}       
\usepackage{graphicx}       
\graphicspath{{media/}}     

\usepackage{cite}
\usepackage{amsmath,amssymb,amsfonts}
\usepackage{algorithmic}
\usepackage{graphicx}
\usepackage{textcomp}
\usepackage{booktabs} 
\usepackage{tabularx} 
\usepackage{xcolor}
\usepackage{ulem}
\usepackage{url}
\usepackage{hyperref}

\usepackage{algorithm}
\usepackage{soul}
\usepackage{amsmath}
\usepackage{subcaption} 
 
\usepackage{amssymb}
\usepackage{pifont}
\title{FedMIL: Federated-Multiple Instance Learning for Video Analysis With Optimized DPP Scheduling
}

\author{
  Ashish Bastola \\
  School of Computing \\
  Clemson University\\
  Clemson, SC, USA\\
  \texttt{abastol@g.clemson.edu } \\
  \AND
  Hao Wang \\
  School of Computing \\
  Clemson University \\
  Clemson, SC, USA\\
  \texttt{hao9@g.clemson.edu} \\
  \And
  Xiwen Chen \\
  School of Computing \\
  Clemson University\\
  Clemson, SC, USA\\
  \texttt{xiwenc@g.clemson.edu}\\
\And
  Abolfazl Razi \\
  School of Computing \\
  Clemson University\\
  Clemson, SC, USA\\
  \texttt{arazi@clemson.edu } \\
}

\begin{document}
\maketitle

\begin{abstract}

Many AI platforms, including traffic monitoring systems, use Federated Learning (FL) for decentralized
sensor data processing for learning-based applications while preserving privacy and ensuring secured information transfer. On the other hand, applying supervised learning to large data samples, like high-resolution images requires intensive human labor to label different parts of a data sample. Multiple Instance Learning (MIL) alleviates this challenge by operating over labels assigned to the 'bag' of instances. 
In this paper, we introduce Federated Multiple-Instance Learning (FedMIL). This framework applies federated learning to boost the training performance in video-based MIL tasks such as vehicle accident detection using distributed CCTV networks. However, data sources in decentralized settings are not typically Independently and Identically Distributed (IID), making client selection imperative to collectively represent the entire dataset with minimal clients. To address this challenge, we propose DPPQ, a framework based on the Determinantal Point Process (DPP) with a quality-based kernel to select clients with the most diverse datasets that achieve better performance compared to both random selection and current DPP-based client selection methods even with less data utilization in the majority of non-IID cases. This offers a significant advantage for deployment on edge devices with limited computational resources, providing a reliable solution for training AI models in massive smart sensor networks. 
\footnote{This material is based upon work supported by the National Science Foundation under Grant Numbers CNS2204721 and CNS-2204445.}

\end{abstract}


\section{Introduction}

In recent years, video-based AI platforms have seen prolific growth and unprecedented applications across diverse sectors. Smart transportation systems, by using Autonomous Vehicles (AVs) and AI-based traffic control platforms, are revolutionizing mobility with enhanced efficiency and safety\cite{alessandrini2015automated, olaverri2016autonomous, coppola2019autonomous, chen2022network, bastola2024driving, bastola2023multi}. 
Specifically, accident detection plays a significant role in ensuring safety during unforeseen circumstances. Prompt detection of accidents enables emergency services to respond on time, while also alerting ongoing traffic to prevent further catastrophic incidents. However, even with millions of decentralized CCTV cameras across the US infrastructures, accident detection remains an often overlooked area \cite{wang2020vision, razi2023deep}.

\begin{figure*}[htbp]
\centering
\includegraphics[width = 0.98\linewidth]{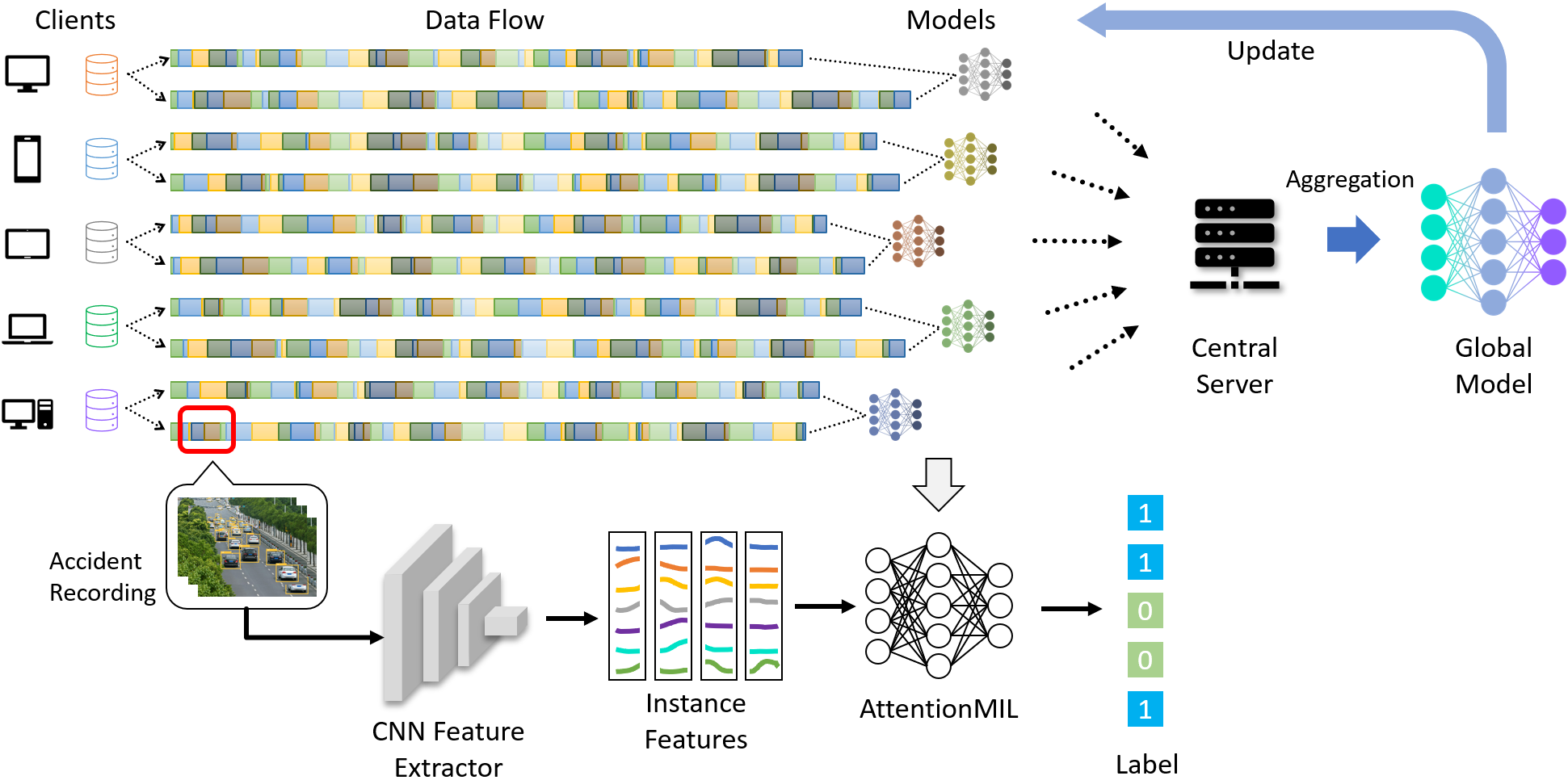}
\caption{The architecture of FedMIL. Compared to classical central learning, the training process is broken down to the client level to prevent massive data exchange. In advance, with a pre-trained CNN model to extract the video feature and a lightweight MIL model for bag-level classification, the computational cost for edge devices is greatly reduced.}
\label{fig:FL}
\end{figure*}

This paper focuses on developing data selection strategies to construct generalizable learning-based models for traffic analysis that require access to a minimal number of roadside units (RSUs) that collectively best represent the entire traffic. Building learning-based inference and control models typically requires access to a sheer number of data sources (roadside cameras, in our case) to collect a massive amount of data to represent various situations. 
Our approach of selective access to the best set of data sources not only minimizes the processing and communication costs but also facilitates the development of reliable models with limited access to a handful of roadside infrastructures.

Centralized data storage and processing usually face several privacy challenges. While training foundational models necessitates extensive data collection of these sensitive data, it is almost impossible to gather this volume of data without potentially infringing upon user privacy with current technologies \cite{li2021survey, yin2021comprehensive}. The risk of data breaches and unauthorized access is a constant threat in such situations \cite{solove2020information, everson2022comparative}. 
Secondly, centralized processing can be inefficient due to the high computational and storage demands placed on central servers, also an issue of latency in data transfer when exchanging a massive amount of data, which prevents the updating capabilities of these systems\cite{konevcny2016federated}.

For these reasons, most AI platforms depart from central learning and bare minimum learning on edge servers towards Federated Learning (FL), with the core idea of sharing local model parameters to implement a global model instead of directly sharing raw data. 
Using this system, each client performs training on its own dataset, and then shares model parameters to build an inclusive global model that can operate in wide geographic areas with varying types, rates, and frequency of crash types.

On the other hand, Multiple Instance Learning (MIL) is a weakly supervised learning paradigm that trains models on bags of data samples that share the same label, which can be applied to various fields such as medical image classification and anomaly detection \cite{zhu2023pdl,qiu2023sc,zhu2023self}. It also has gained widespread use in video-based incident detection applications by treating each video frame as an individual instance and the entire video as a bag of such instances \cite{BaoMM2020, lv2023unbiased}. Furthermore, efforts have been made to enhance these models by focusing on multi-scale continuity \cite{gong2022multi}. This enhancement is based on the understanding that anomalous events in real-world videos typically unfold continuously, necessitating their evaluation across the entire bag of instances.
The use of multiple instance learning significantly reduces the need for extensive data preprocessing and feature engineering, which reduces both the labor cost and the demand for computational resources to train AI models on edge devices.

To combine both advantages, we propose FedMIL, the federated version of MIL, as a technical contribution.
As shown in Figure \ref{fig:FL}, MIL provides the benefit of developing weakly supervised models to reduce the computation cost on the client side, and FL provides the benefit of avoiding massive data exchange. 
Different than another federated MIL setting that deploys models on only a few sites for weakly-supervised image classification in \cite{lu2022federated}, our study focuses on the video-based analysis among hundreds of data nodes, where the total data flow and computational cost is thousands of times than image-based tasks. Furthermore, compared to image classification tasks, our proposed framework can be extended to other tasks that are more practical, such as video anomaly detection, traffic volume analysis, human action recognition, and deepfake detection \cite{lv2023unbiased, chen2022network, li2020sharp}.

Although the combination of FL and MIL eliminates the need for massive data exchange and reduces computation power requirements for edge devices for complex tasks, orchestrating model parameter exchange among a sheer number of clients might not be feasible for practical reasons. 
For instance, the FL scheduler system only allows a limited number of communication ports at one time, and the network bandwidth is highly constrained. 
Meanwhile, due to the inherent diversity of data across different clients \cite{zhu2021federated}, the rate and types of crashes can be extremely different across different cameras in the real world. 
For instance, an RSU located in a less crowded area or a safe zone might not witness frequent accidents, thus the locally trained model may fail to detect accidents if one occurs \cite{chen2022network}. 
Thus, these non-IID data distributions may greatly impact the model's performance due to weight divergence issues \cite{luping2019cmfl, cho2020bandit}.

Therefore, developing client selection strategies to reduce the negative impact of uneven data distribution, and improve the performance under constrained communication or limited local processing power becomes a key challenge in this research area.

Intuitively, selecting RSUs with more diversity in terms of data and label distributions is more desirable.
In recent studies, the Determinantal Point Process (DPP) achieved significant gains in federated learning applications for capturing diversity \cite{kulesza2011learning}.
DPP enforces repulsion to alike items and prioritizes selecting clients whose data distributions are varied, aiming to reduce redundancy in the training process. With classical DPP-based client selection, the central model tends to collect model parameters from clients with more diverse data, avoiding convergence at lower accuracy \cite{zhang2023dpp}.

We take this step further by proposing a power-of-choice \cite{cho2020client} version of DPP embedded as a quality matrix to achieve the best of both diversity and loss gradients, which makes the selection even more robust\cite{cho2020client}. 
Thus, the second contribution of this work is the demonstration of the efficacy of selecting a diverse subset of clients that accurately represent the overall dataset, which ensures the results are robust and broadly applicable across various clients in real-world settings.

\section{Problem Formulation}

The key aspect of the problem in this study is the data distribution among clients. While IID sampling offers a baseline approach where each client receives equitable and balanced representative samples, non-IID sampling models the skewness in data distribution, which aligns more closely with real-world data distribution scenarios\cite{li2022federated, lv2023unbiased} 

\begin{figure*}[htbp]
\centering
\includegraphics[width=0.98\linewidth]{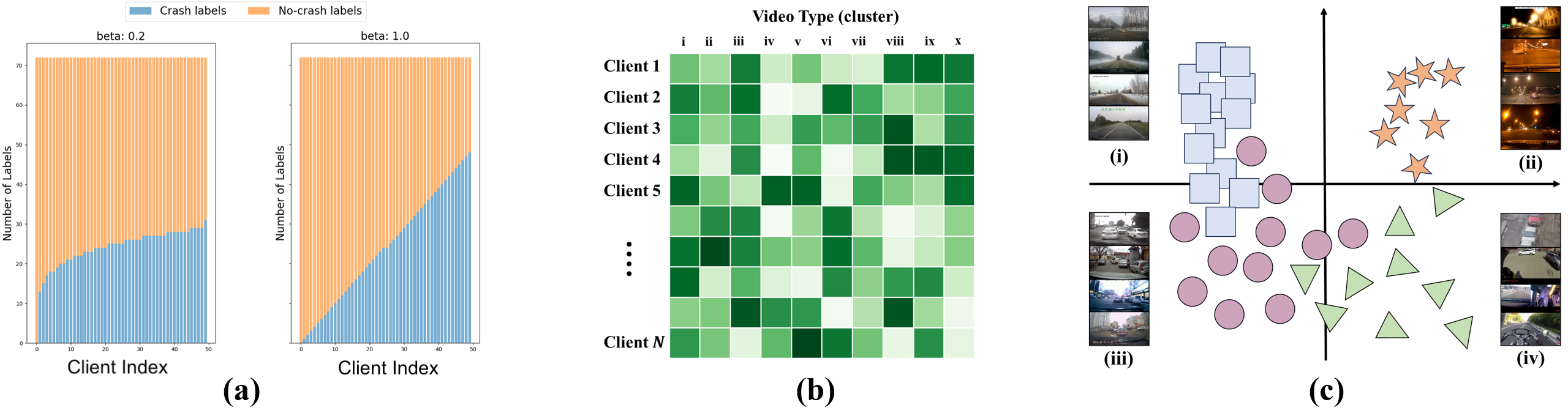}
\caption{Simulating real-world data distribution: (a) the data distribution of different clients is imbalanced directly at the class level; (b) data imbalance based on Dirichlet distribution over underlying data clusters, and (c) K-means clustering on VGG features to identify underlying clusters, where (i) represents the video condition under snow day, (ii) represents the night snippets, (iii) is the urban recording, and (iv) represent other scenarios.}
\label{fig:data_dist}
\end{figure*}

To simulate non-IID distribution, we introduce class and feature-space-based imbalance, as illustrated in Figure \ref{fig:data_dist}. 
For the first type, we use power-law to assign imbalanced crash labels to each client \cite{li2019convergence, balakrishnan2022diverse, jamali2022federated}. 
This mirrors the situations where some RSUs may observe crashes or specific types of crashes more frequently than others. 
For the second type, we introduce an underlying distribution-based imbalance such that each client receives a subset whose data type is imbalanced according to a Dirichlet distribution \cite{li2022federated}. 
For instance, the traffic compositing on a highway contains more trucks than at a city intersection. Therefore, their features extracted from video frames may follow statistically different distributions.

Specifically, if labeled video pairs are shown by $(V_i,y_i)=([F_i(1),F_i(2),\cdots,F_i(M)],y_i)$ with $V_i$ being a video clip, $F_i(j)$ being its $j^{th}$ frame, and $y_i$, being the class label, then we impose imbalance to both $V_i$ and $y_i$.We apply power-law to class labels $y_i$ and Dirichlet distribution to samples $V_i$ as detailed next, to divide the data into non-IID subsets among clients.

\subsection{Label-based Imbalance (Type I Non-IID)}
We follow the imbalance strategy for the binary distribution by changing the ratio of class labels for each client. We vary the ratio of class labels client-wise by tuning the power-law. 

Specifically, we use:
\begin{equation}
\label{eq:power_law}
n_p=V \cdot(p+H)^\beta
\end{equation}
where $n_p$ is the proportion of class 1 counts for client $p$, $V$ and $H$ denote the vertical and horizontal scaling, respectively, and the exponent $\beta$ is a tuning factor to control the 
the ratio of class labels. 
We fix the horizontal and vertical scaling parameters to certain values to account for the unequal count of either of the classes before beginning the distribution, then vary $\beta$ to generate different levels of disparity. To add non-uniformity in label assignment, we select a hold-out set from both classes in the two selected distributions of $\beta$, so some clients are assigned only one label type. Here, we assume binary labels for crash detection, but it easily extends to multi-level cases.

\subsection{Distribution-based Imbalance (Type II Non-IID)}
While class-based imbalance provides disparity of labels, the distribution-based imbalance better captures the disparity among the attributes of video frames that represent real-world factors, such as climate conditions (sunny, snowy, and cloudy), road features (off-road, highway, or regular), environment (urban and rural), light conditions, etc.  

The process includes the following steps. Assume a dataset offers video clips $V_i$ and a set of feature vectors $\phi_W(F_j)$ for each frame $F_j$ extracted using a pre-trained VGG network. These feature vectors are clustered using k-means \cite{hartigan1979algorithm} to identify $C=10$ different underlying classes as shown in Fig. \ref{fig:data_dist}. Then, we perform majority voting on the obtained feature vectors to label the sample (aka video clip $V_i$). Finally, we apply Dirichlet distribution to create the imbalance corresponding to these cluster labels. We follow the strength variation of this distribution as followed by \cite{li2022federated}, where we sample $\vec{n}_p=[n_p^1,n_p^2,\cdot,n_p^C] \sim \operatorname{Dir}(\vec{n};\vec{\alpha})$ using
\begin{equation}
\small
    \operatorname{Dir}(\vec{n} ;\vec{\alpha})= 
    \frac1{B(\vec{\alpha})}\prod_{c=1}^C n_p^{\alpha_c-1}
\end{equation}
and allocate $n_p^c$  proportion of the instances of class $c$ to client $p$, where $\operatorname{Dir}(\cdot)$ denotes the Dirichlet distribution and $\vec{\alpha}=[\alpha_1,\alpha_2,\cdot, \alpha_C]$ denotes the concentration parameter ($\alpha_c > 0$). Here, $\mathrm{B}(\vec{\alpha})$ is the multinomial beta function given by: 

\begin{equation}
\label{eq:labeldir}
\small
   \mathrm{B}(\vec{\alpha})=\frac{\prod_{c=1}^C\Gamma(\alpha_c)}{\Gamma\left(\sum_{c=1}^C\alpha_c\right)} 
\end{equation}

where $\text{L}$ is the gamma function given by $\Gamma(\alpha_c)=(\alpha_c-1)!$. 
The major advantage of this approach is that we can change the imbalance level by varying the concentration parameter $\alpha$. The smaller the $\alpha$, the more unbalanced the partition.

\section{Algorithm Design}
Our contribution can be divided into two folds.
First, we optimized the existing FL framework to better serve the MIL specification. In our case, the FL architecture is specifically designed for video analysis, where each frame (instance) has multiple features.
Second, we propose the client selection algorithm based on the DPPQ method.
Proposed by \cite{kulesza2011learning}, the quality-diversity decomposition is the special reparameterization of the normal DPP-based method. 
This method has been proven to achieve a significant gain in video summarization applications \cite{gong2014diverse}. We show that it can be applied to similar tasks, such as the video classification in this work.

\subsection{Federated Learning Framework }

\subsubsection{Individual Client Model and Training}
Each client $p$ trains the model on their local data $D_p$ by minimizing the local loss function $L_p(w)$ using stochastic gradient descent:
\begin{equation}
\small
\label{eq:local_objective}
    L_p = \frac{1}{|D_p|} \sum_{k=1}^{|D_p|} L_k(g_w(\vec{z}_k), \vec{y}_{k})
\end{equation}
where $L_p$ is the loss function for each client $p$. $\vec{y}_k$ is the label, $\vec{z}_k$ is the bag level representation(vector in our case) and $L_k$ is the loss for each sample $k$, and $g_w$ is the model parameterized by weights $w$.

\subsubsection{Global Aggregation}
After local training, clients send their model weights $w_p$ to the central server. The server performs a weighted average to update the global model:
\begin{equation}
\small
\label{eq:global_aggregation}
    w_g^{t+1} = \sum_{p=1}^{P} \frac{n_p}{n} w_p^t
\end{equation}
where, $w_g^{t+1}$ is the aggregated global model ready for $t+1$-th round, $n_p$ is the number of samples at client $p$, $n$ is the total number of samples, and $t$ denotes the current round.

\subsection{Model Design for FedMIL}
We apply AttentionMIL \cite{ilse2018attention}, A multi-instance learning model that processes bags of instances. It assigns attention weights to instances, enabling the model to focus on the most informative instances in each bag. 

Thus, for the weighted average of instances where the image feature determines the weights,
the extracted embeddings $\mathbf{H}=\left\{\mathbf{h}_1, \ldots, \mathbf{h}_K\right\}$ are the bag of $K$ embeddings for $K$ samples(which have n instances each) that are the output of the feature extractor (which is a Fully connected layer). Thus, as proposed by \cite{ilse2018attention, dauphin2017language}, the MIL Pooling or the bag-level aggregation to form the bag representation for the kth sample is given by:

\begin{equation}
\label{eq:z_k}
\small
\vec{z}_k=\sum_{m=1}^n a_m \vec{{h}}_m
\end{equation}
where $a_m$ is the gated attention with $\tanh(\cdot)$ non-linearity given by, 

\begin{equation}
\small
a_m=\frac{\exp \left\{\mathbf{w}^{\top}\left(\tanh \left(\mathbf{V} \vec{h}_m^{\top}\right) \odot \operatorname{sigm}\left(\mathbf{U }\vec{h}_m^{\top}\right)\right)\right\}}{\sum_{j=1}^n \exp \left\{\mathbf{w}^{\top}\left(\tanh \left(\mathbf{V}\vec{h}_j^{\top}\right) \odot \operatorname{sigm}\left(\mathbf{U }\vec{h}_j^{\top}\right)\right)\right\}}
\end{equation}

where $\mathbf{U} \in \mathbb{R}^{L \times M}$ are parameters, $\odot$ is an element-wise multiplication and $\operatorname{sigm}(\cdot)$ is the sigmoid non-linearity. The gating mechanism introduces a learnable non-linearity that potentially removes the troublesome linearity in $\tanh (\cdot)$. 

The Attention-based MIL handles the multi-instance nature of the video data. We utilize the stochastic gradient descent (SGD) optimizer with a Lookahead mechanism for robust optimization \cite{zhang2019lookahead}. The model was trained using a cross-entropy loss function suitable for the binary classification task (accident vs. non-accident). Thus, the global loss function is given by:


\begin{equation}
\small
    L_g=\sum_{p=1}^P \frac{n_p}{\sum_{p \in \mathcal{P}} n_p} L_p
\end{equation}

where $P$ is the total number of selected clients, $L_p$ is the loss for $p$-th client defined in \ref{eq:local_objective} (however $\vec{z}_k$ now is the attention-based bag-level aggregation defined in \ref{eq:z_k} for $k$-th bag). Based on this, we formulate the Federated learning architecture to train multiple clients in a decentralized fashion.

\subsection{Client Profiling}
Assume each sample of the dataset is represented as an image feature. Considering the $p$-th client, let the $k$-th sample be represented as $x_k$. Each sample contains 50 features corresponding to the image frame. Then, the $i$-th feature corresponding to $i$-th frame undergoes the following feature extraction $\vec{h}_i=\text{ReLU}(\mathbf{W}\mathbf{x}_i+\mathbf{b})$. Note that this intermediary feature extractor helps further ensure security, which makes them robust to inference-based attacks. Thus, for the $k$-th sample, we have $\mathbf{h_k}=\left\{\vec{h}_1, \ldots, \vec{h}_n\right\}$ where $n=50$ is the total number of frames in a single video sample. Since we employ SGD, the batch size is the total number of samples for each client after the distribution. Thus for each $p$th client with dataset $D_p$, we compute the average feature profile $\mathbf{h_p}$ given by:
\begin{equation}
\small
\label{eq:hp}
\mathbf{h_p}=\frac{1}{|D_p|}\sum_{k=1}^{|D_p|} \mathbf{h}_k
\end{equation}

Further, we profile the average loss of each client as follows, 
\begin{equation}
\small
\label{eq:lp}
L_p=\frac{1}{|D_p|}\sum_{k=1}^{|D_p|} {L}_k
\end{equation}
where $L_p$ is the loss profile for the $p$-th client and $L_k$ is the loss value computed for each sample $k$.

\subsection{Client Selection via Quality-Diversity Decomposition}

Let $\mathcal{Y}=\{1,2, \cdots, N\}$ be the ground set of N clients. We then profile each $p$-th client ${C_p}$ using the average output of their feature representation given by Eq. \ref{eq:hp} and their average loss given by Eq. \ref{eq:lp}, which are the only data shared to the central server as their representative information and only during the initialization phase.

With this information, the central server first computes the similarity matrix $\mathbf{S}=\{s_{r,t}\}_{C\times C}$ as implemented in \cite{zhang2023dpp, kulesza2011learning} using the following:

\begin{equation}
\small
\label{eq:sim_matrix}
    s_{r,t}=1-\left(\frac{s_{r,t}^0-\min(\mathbf{S}^0)}{\max(\mathbf{S}^0)-\min(\mathbf{S}^0)}\right)
\end{equation}

where, $s_{r,t}^0\text{ = }\left\|\mathbf{h}_r-\mathbf{h}_t\right\|_{2} $ with $\mathbf{h}_r,\mathbf{h}_t \forall {r, t} {\operatorname*{\in}}{C}$, and $\mathbf{S^0}$ is the full difference matrix of every client combination defined by $\{s_{r,t}^{0}\}_{C\times C}$.


Similarly, the quality matrix is defined as the diagonal matrix of loss values $\mathbf{L_p}$ pertaining to each client $C_p \in \mathcal{Y}$ i.e. 

\begin{equation}
\small
\label{eq:quality_matrix}
    \mathbf{Q}=\begin{pmatrix}q_1&0&\cdots&0\\0&q_2&\cdots&0\\\vdots&\vdots&\ddots&\vdots\\0&0&\cdots&q_N\end{pmatrix}
\end{equation}
where, $q_p$ for $p \in \{1,2, \cdots, N\}$ is the min-max scaled quality value of the $p$-th client with lowerbound $\epsilon$ given by eq \ref{minmax}. This is important to prevent a loss value of 0 that ruins the positive semidefinite property of the kernel matrix.

\begin{equation}
\small
\label{minmax}
    q_{\mathrm{p}}=\epsilon+\left(\frac{{L_p}-{L_{\min}}}{{L_{\max}}-{L_{\min}}}\times(1-\epsilon)\right)
\end{equation}

We finally calculate the kernel matrix given by $\mathbf{L}=\mathbf{Q}\mathbf{S}^\mathrm{T}\mathbf{S}\mathbf{Q}$. And following \cite{kulesza2011learning}, we can express the probability of selecting subset $G$ from $\mathcal{Y}$ as, 

\begin{equation}
\small
\label{eq:PL}
    P_L(G\subseteq \mathcal{Y}) \propto det(\mathbf{L_G}),
\end{equation}

Thus, from Eq. \ref{eq:PL}, we can infer the select clients with higher diversity-quality score.
i.e. the loss value of the client $L_p$ and the diversity of the extracted features $\mathbf{h_k}$. Thus our method incorporates a selection of clients with higher loss values and are similarly highly diverse. Here, we sample $P$ clients and denote their index set as $G$. The more detailed working of the algorithm is mentioned in Algorithm 1.

\begin{algorithm}[ht]
\small
\label{alg:1}
\caption{FL-DPPQ: Federated Learning with DPP-based Quality Diversity Decomposition} 
\begin{algorithmic}
\REQUIRE Clients with global model parameters $\mathbf{w}_g^{(0)}$ having their own data distributed using some non-IID distributions defined in Eq. \ref{eq:labeldir} and \ref{eq:power_law}

\FOR {each client $c$ $\in \mathcal{Y}$}
    \item Calculate the average loss and average of extracted features using Eqs. \ref{eq:lp} and \ref{eq:hp} and upload it to the server.
\ENDFOR
\item The Central Server calculates the similarity matrix S according to Eq. \ref{eq:sim_matrix} and quality matrix Q according to \ref{eq:quality_matrix} and constructs $k$-DPPQ and selects $G$ of $\mathcal{Y}$ clients with probability given by Eq. \ref{eq:PL}.
\FOR {$t$ \textbf{in} $1:T$} 
    \FOR {each client $c$ $\in G$} 
        \FOR {each local epoch $t'$ $\in 1:T'$} 
            \item Update $\mathbf{w}_c^{(t)}$ using back prop. while minimizing the local objective function given by Eq. \ref{eq:local_objective}.
        \ENDFOR
        \item Send updated weight $\mathbf{w}_c^{(t)}$ to the server.
    \ENDFOR
    \item Central server performs the weight aggregation using Eq. \ref{eq:global_aggregation} and distributes the updated global model to all the clients.
\ENDFOR
\ENSURE The trained Global Model $\mathbf{w}_g^{(T)}$.
\end{algorithmic}
\end{algorithm}

\section{Experiments }

We show the proposed work on CCD, a typical multi-instance annotated dataset, and we compare the model performance under different settings (e.g., data utilization rate, strength of data imbalance, and non-IID type). 
The results are averaged over 20 runs. The model was evaluated on a test set with performance metrics including loss, accuracy, F1 score, and Area Under the Receiver Operating Characteristic Curve (AUC-ROC) \cite{kamoona2023multiple}. 
The result shows that model performance varies in terms of data distribution type. Moreover, our implementation of FedMIL improved the model performance by selecting quality clients via the proposed DPPQ, especially when the data utilization decreased. 


\subsection{Car Crash Dataset (CCD) for Traffic Accident Analysis}

The Car Crash Dataset (CCD) is a specialized collection for traffic accident analysis. It comprises 1,500 dashcam-captured accident videos and 3,000 normal videos from YouTube and the BDD100K dataset, each containing 50 frames at 10 fps \cite{BaoMM2020}. Advanced feature extraction is employed for all videos using VGG-16 for frames and bounding boxes, detected via Cascade R-CNN with a ResNeXt-101 backbone. Thus, each video is annotated with frame-level binary labels (crash, no crash)

\subsection{Platform and Training Process}
All experiments are trained on the Linux cluster server with AMD CPU and NVIDIA GPU.  A typical configuration for each run is: 8 CPU Core, 15GB RAM, and 1$\times$A100 GPU. 
For each hyper-parameter test, the result is based on the average of 20 individual runs. Each individual run includes 50 epochs of training to reach the converge point.

\subsection{Evaluation on MNIST Dataset}
\begin{figure}[htbp]
 \centering
 \includegraphics[width=0.5\linewidth]{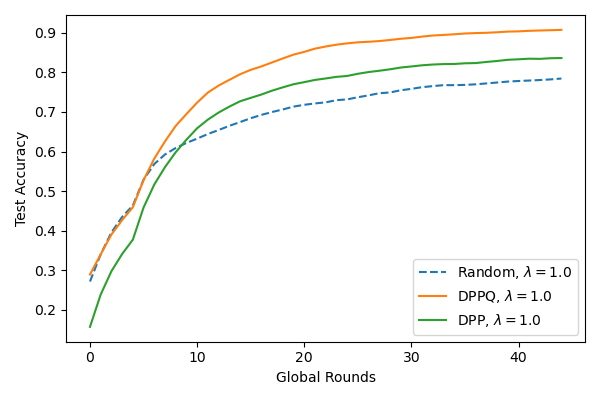}
 \caption{Training comparison in MNIST dataset with Type I Non-IID distribution of strength $(\alpha=0.5)$ and 100\% data utilization}
 \label{fig:mnist}
 \end{figure}

To validate the performance of the proposed DPPQ method in general, we test our method on the MNIST dataset that has 10 different classes. We perform an average of 10 runs for extreme non-IID data distribution with Type I Non-IID distribution of strength $(\alpha=0.5)$ and full data utilization $(\lambda=1.0)$. 
As shown in table \ref{tab:mnist}, the proposed DPPQ outperforms both random selection and DPP with a considerable gain. Furthermore, Figure \ref{fig:mnist} shows that DPPQ can reach a higher accuracy in the training process, which demonstrates the benefit of client selection with quality consideration.

 \begin{table}[htpb]
 \centering
 \caption{Model Comparison in MINIST}
 \label{tab:mnist}
 \resizebox{0.2\textwidth}{!}{%
 \begin{tabular}{@{}ll}
 \toprule
 Category & Test Accuracy\\ \midrule
 Random            & 0.7869\\
 DPP               & \multicolumn{1}{l}{0.8370}  \\
 DPPQ              & \multicolumn{1}{l}{\textbf{0.9084 (+8.5\%)}   } \\
 \bottomrule
 \end{tabular}%
 }
 \end{table}

\subsection{Type I Non-IID Evaluation}
\begin{figure*}[htbp]
\centering
\includegraphics[width=0.9\linewidth]{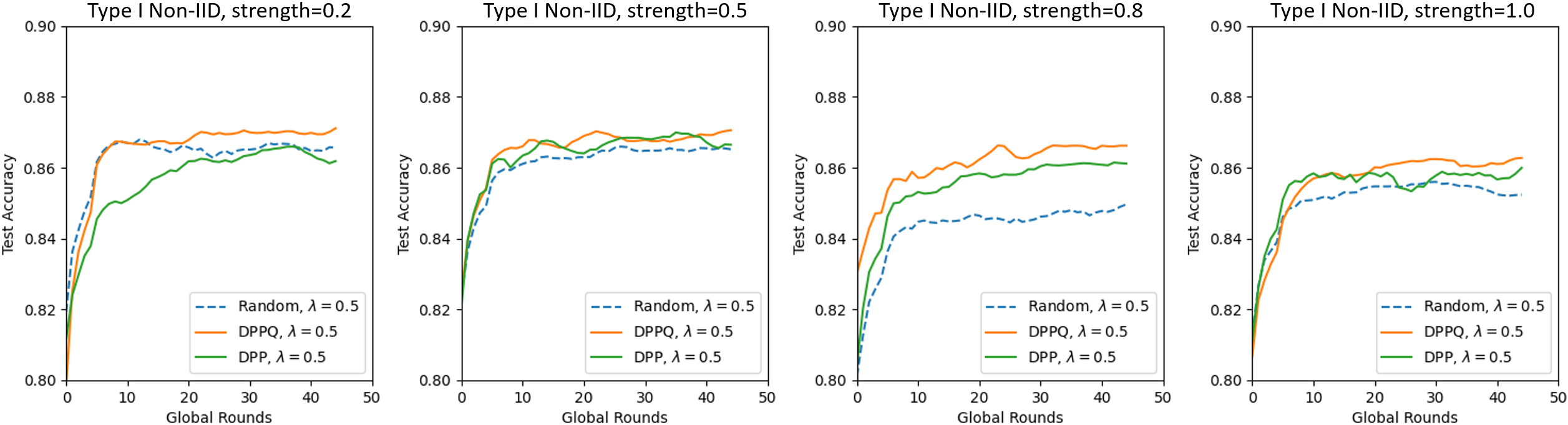}
\caption{Model performance comparison in terms of data imbalance based on class labels.}
\label{fig:iid2_1}
\end{figure*}

Figure \ref{fig:iid2_1} shows the comparison between simple DPP-based sampling as presented in \cite{zhang2023dpp}, random and our DPPQ-based sampling in the CCD dataset. The different strength values in this case are the tuning parameter for $\beta$ as defined in eq. \ref{eq:power_law} with the data utilization of $50\%$. From the results, we can infer that DPPQ achieved faster convergence than DPP when analyzed over a strength value of $0.2$. We also see random performs faster but fails to maintain the same accuracy as DPPQ after the 15th epoch. We observed similar faster convergence in strength value of $0.8$, achieving nearly $2\%$ accuracy gain compared to random selection.

\begin{table*}[htbp]
\centering
\caption{Type I Non-IID Model Comparison}
\label{tab:type2}
\resizebox{0.9\textwidth}{!}{%
\begin{tabular}{cc|lll|lcc|lcc|lll}  
 & &   \multicolumn{3}{c|}{Strength = 0.2}&\multicolumn{3}{c|}{Strength = 0.5}&  \multicolumn{3}{c}{Strength = 0.8}& \multicolumn{3}{|c}{Strength = 1.0}\\ \toprule
 Data Utilization& Method&    aucs&f1& test acc& aucs&f1&test acc&  aucs&f1& test acc & aucs& f1&test acc\\
  \toprule
 0.5& DPPQ
&    \textbf{0.9443}&\textbf{0.8689}& \textbf{0.8703}& 0.9411&\textbf{0.8675}&\textbf{0.8700}&  \textbf{0.9425}&\textbf{0.8645}& \textbf{0.8658}& 0.9353& \textbf{0.8614}&\textbf{0.8621}
\\
 & DPP
&    0.9394&0.8619& 0.8632& 0.9425&0.8660&0.8676&  0.9425&0.8638& 0.8624& 0.9382& 0.8581&0.8589
\\
 & Random
&   0.9404&0.8649& 0.8657& 0.9386&0.8626&0.8632&  0.9277&0.8473& 0.8485& 0.9335& 0.8507&0.8463
\\
\midrule
 0.8& DPPQ
&    0.9449&\textbf{0.8738}& \textbf{0.8773}& 0.9535&0.8794&0.8812&  0.9505&\textbf{0.8799}& \textbf{0.8791}& \textbf{0.9511}& \textbf{0.8777}&\textbf{0.8765}
\\
 & DPP
&    0.9432&0.8684& 0.8720& 0.9524&0.8770&0.8778&  0.9576&0.8730& 0.8712& 0.9462& 0.8700&0.8695
\\
 & Random
&    0.9503&0.8736& 0.8735& 0.9578&0.8884&0.8880&  0.9441&0.8652& 0.8652& 0.9488& 0.8755&0.8754
\\ 
\midrule
 1.0& DPPQ
& 0.9564& \textbf{0.8859}& \textbf{0.8835}& 0.9568& 0.8851& 0.8854& 0.9508& 0.8806& 0.8791& 0.9508& 0.8779&\textbf{0.8773}\\
 & DPP
& 0.9579& 0.8843& 0.8829& 0.9585& 0.8848& 0.8861& 0.9562& 0.8833& 0.8831& 0.9523& 0.8784&0.8768\\
 & Random
& 0.9562& 0.8829& 0.8822& 0.9584& 0.8918& 0.8935& 0.9518& 0.8733& 0.8717& 0.9517& 0.8756&0.8726
\\
\bottomrule
\end{tabular}%
}
\end{table*}

\subsection{Type II Non-IID Evaluation}

\begin{figure*}[htbp]
\centering
\includegraphics[width=0.9\linewidth]{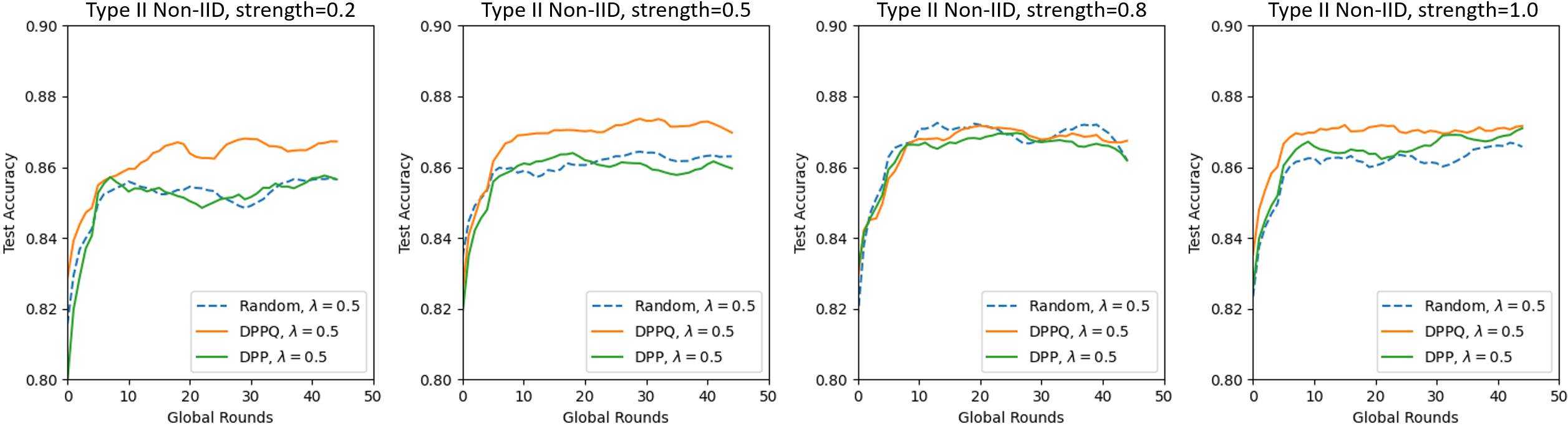}
\caption{Model performance comparison in terms of data imbalance based on Dirichlet distribution.}
\label{fig:iid3_3}
\end{figure*}

Figure \ref{fig:iid3_3} shows the comparison between random selection, DPP, and the proposed DPPQ method in the CCD dataset under the Dirichlet distribution with a data utilization rate of $ 50\%$, where the strength represents the data imbalance ratio. In advance, $\alpha=0.2$  represents extremely imbalanced, $\alpha=0.5$ represents moderately imbalanced, $\alpha=0.8$  represents almost balanced, and $\alpha=1.0$  indicates the data volume for each client is almost even.

As a result, the overall performances of all models are dropped due to the lower data utilization. Nevertheless, DPPQ outperforms DPP and random selection regardless of the data imbalance ratio. This is a significant advantage when the client's device is busy or has limited computation resources.

\begin{table*}[htbp]
\centering
\caption{Type II Non-IID Model Comparison}
\label{tab:type3}
\resizebox{0.9\textwidth}{!}{%
\begin{tabular}{cc|lll|lcc|lcc|lll}  
 & &   \multicolumn{3}{c|}{Strength = 0.2}&\multicolumn{3}{c|}{Strength = 0.5}&  \multicolumn{3}{c}{Strength = 0.8}& \multicolumn{3}{|c}{Strength = 1.0}\\ \toprule
 Data Utilization& Method&    aucs&f1& test acc& aucs&f1&test acc&  aucs&f1& test acc & aucs& f1&test acc\\
  \toprule
 0.5& DPPQ
&    \textbf{0.9389}&\textbf{0.8630}& \textbf{0.8660}& \textbf{0.9411}&\textbf{0.8676}&\textbf{0.8713}&  0.9439&0.8669& \textbf{0.8683}& \textbf{0.9444}& \textbf{0.8703}&\textbf{0.8710}\\
 & DPP
&    0.9316&0.8538& 0.8561& 0.9378&0.8604&0.8596&  0.9465&0.8538& 0.8643& 0.9437& 0.8683&0.8693
\\
 & Random
&   0.9276&0.8531& 0.8563& 0.9378&0.8579&0.8630&  0.9436&0.8685& 0.8670& 0.9374& 0.8628&0.8658
\\
\midrule
 0.8& DPPQ
&    0.9475&\textbf{0.8719}& \textbf{0.8733}& 0.9532&\textbf{0.8789}&\textbf{0.8782}&  0.9563&\textbf{0.8837}& \textbf{0.8855}& \textbf{0.9525}& \textbf{0.8782}&\textbf{0.8792}
\\
 & DPP
&    0.9485&0.8705& 0.8705& 0.9534&0.8786&0.8776&  0.9574&0.8802& 0.8809& 0.9525& 0.8779&0.8785
\\
 & Random
&    0.9413&0.8603& 0.8589& 0.9510&0.8753&0.8761&  0.9539&0.8825& 0.8820& 0.9511& 0.8751&0.8765
\\ 
\midrule
 1.0& DPPQ
& \textbf{0.9581}& \textbf{0.8857}& \textbf{0.8860}& 0.9564& \textbf{0.8870}& \textbf{0.8870}& 0.9527& 0.8815& 0.8842& 0.9575& 0.8871&\textbf{0.8880}\\
 & DPP
& 0.9531& 0.8816& 0.8830& 0.9568& 0.8851& 0.8858& 0.9586& 0.8853& 0.8856& 0.9602& 0.8878&0.8880\\
 & Random
& 0.9465& 0.8710& 0.8694& 0.9543& 0.8844& 0.8856& 0.9560& 0.8869& 0.8874& 0.9544& 0.8825&0.8844
\\
\bottomrule
\end{tabular}%
}
\end{table*}

Table\ref{tab:type3} on the other hand, shows the complete results of all experiments. 
As the data utilization is controlled to 100\%, DPPQ shows identical performance to the DPP and random selection when Dirichlet strength $\alpha$ is $0.5$ and $1.0$, this is because the full-utilized client data enhanced the overall diversity, and the reduction of data imbalance (increase of $\alpha$) will improve all model performance, especially the random selection.
However, when data become more imbalanced ($\alpha=0.2$), the proposed DPPQ shows better performance than DPP and random selection. This is because the proposed DPPQ not only considered the data volume but also the data quality of each client. 
With the data utilization dropped to 80\%, the overall performances of all models are dropped due to the lower data utilization. Nevertheless, the proposed DPPQ and DPP still retain a better performance than the random selection.

\subsection{Data Utilization Analysis}

\begin{figure}[htbp]
\centering
\includegraphics[width=0.5\columnwidth]{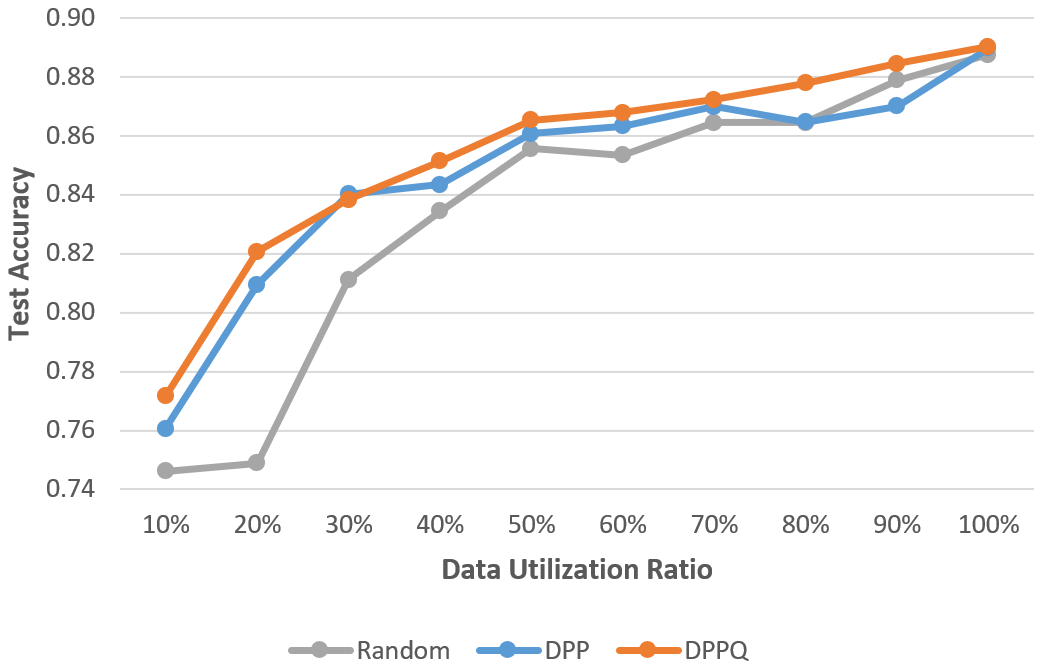}
\caption{Model performance comparison in terms of data utilization ratio.}
\label{fig:data_rate}
\end{figure}

Figure \ref{fig:data_rate}, on the other hand, shows the model performance in terms of data utilization ratio. The result is using type-II non-IID (Dirichlet distribution) with strength $\alpha=0.5$. Since the proposed DPPQ is optimized for Dirichlet distribution, its performance is better and more stable than DPP and random selection. As the data utilization keeps dropping, the proposed DPP still outperforms random selection, proving its ability for low data-flow scenarios.

\section{Conclusion}
In this study, we presented an approach for accident detection based on traffic videos, integrating a federated learning framework with multiple instance learning to boost the training process on edge devices. Furthermore, we introduce an improved client selection strategy based on DPPQ, which prioritizes diversity and loss gradient. 
This strategy not only enhances the model's generalization capabilities but also provides robust performance under low-data utilization compared to the classical DPP-based approach, which we also validated in the MNIST dataset as a general benchmark.
The test results of the CCD dataset prove that the proposed DPPQ model retains identical performance even though the data become extremely imbalanced. More importantly, the proposed DPPQ model performs better than classical DPP and random selection even with less data utilization.
Our implementation addresses the challenges of both non-IID data distributions and limited computation power in practical applications in the real world.

In advance, the proposed FedMIL framework can be easily extended to more practical video-based tasks such as video anomaly detection, traffic volume analysis, human action recognition, and even deepfake detection, with ensured privacy and security.

\section*{Acknowledgements} 
This material is based upon work supported by the National Science Foundation under Grant Numbers CNS2204721 and CNS-2204445.

\bibliography{references} 
\bibliographystyle{unsrt}

\end{document}